\documentclass[12pt]{spieman}
\usepackage{graphicx}
\usepackage{setspace}
\usepackage{amsmath,amsfonts,amssymb}
\usepackage{tocloft}

\title{Coupled acoustoplasmonic resonators: the role of geometrical symmetries }

\author[a]{Beatriz Castillo L\'opez de Larrinzar}
\author[a]{Jorge M. Garc\'ia}
\author[b]{C. Xiang}
\author[b,+]{N. D. Lanzillotti-Kimura}
\author[a,*]{Antonio Garc\'ia-Mart\'in}
\affil[a]{Instituto de Micro y Nanotecnolog\'ia IMN-CNM, CSIC, CEI UAM+CSIC, Isaac Newton 8, Tres Cantos, 28760 Madrid, Spain}
\affil[b]{Université Paris-Saclay, CNRS, Centre de Nanosciences et de Nanotechnologies, 
10 Boulevard Thomas Gobert, Palaiseau 91120, France}

\cftpagenumbersoff{figure}
\cftpagenumbersoff{table} 
\begin{document} 
\maketitle

\begin{abstract}

Acoustoplasmonic resonators, such as nanobars and crosses, are efficient acousto-optical transducers. The excitation of mechanical modes in these structures strongly depends on the spatial profile of the eigenmodes of the resonator. Using a system of two identical gold elongated bars placed on a silicon dioxide substrate, we examine how breaking mirror symmetries affects the optical and acoustic properties to provide insights in the design of acoustoplasmonic metasurfaces for nonsymmetric acousto-optical transducers. Our findings show that, the absence of mirror symmetries affects differently the optical and nanomechanical response. Broken mirror symmetries not only couple nanomechanical modes existing in single bars, but introduces new torsional resonant modes.
\end{abstract}


{\noindent \footnotesize\textbf{$^+$}N.D. Lanzillotti-Kimura,  \linkable{daniel.kimura@c2n.upsaclay.fr} }
{\noindent \footnotesize\textbf{*}A. Garc\'ia-Mart\'in,  \linkable{a.garcia.martin@csic.es} }
\begin{spacing}{2}   

\section{Introduction}
\label{sect:intro}  

Symmetries are paramount to understand the physical properties of a given system\cite{sep-symmetry-breaking}. 
These symmetries span to all magnitudes, as they impose restrictions, e.g. time reversal invariance makes the scattering matrix symmetric, or the absence of absorption makes it unitary. Others are ``simply'' related to the geometrical aspects of the nanostructures. In many cases the physics is determined by the presence of symmetries, in many others the interest lies in breaking them. A beautiful example is provided by chiral systems. \cite{IsaMarzan,IsaMarzan2,Bea_nanomaterials}

In optics, simple geometries, when brought into close contact, might reveal scattering processes that are highly nontrivial due to the break of symmetry, e.g. chiral behavior of non-chiral interacting elements \cite{stackedHan, BeaAcustoPlasm}. 

In nanoacoustics,~\cite{APL.Perspectives} most of the reported results are based on simple structures, or metasurfaces which present a high degree of spatial symmetry.  For example, bars, crosses, arrays, micropillars, and microdisks all present mirror symmetry planes.\cite{NDLK_Metasurfaces,DellaPicca2016,O’Brien2014_NatComm, APL_Ortiz_10.1063/5.0026959,  Favero_PhysRevLett.105.263903, cavities_Armani2003}  Even the engineering of more complex devices, such as directional acoustic nanoantennas, presents these symmetry planes.  \cite{cubitos,DellaPicca2016} 

It is expected that breaking these symmetry planes would open a plethora of different phenomena to tailor the fundamental modes in nanoacoustic resonators. For example, two concatenated superlattices presenting inverted symmetries \cite{optica_Ortiz:21} give rise to a topologically protected nanoacoustic mode. However, despite recent efforts,~\cite{BeaAcustoPlasm} how broken symmetries enable nanoacoustic modes in three-dimensional nanostructures remain an open question. 

In this work, we make use of numerical simulations to theoretically explore the idea of breaking spatial symmetries to engineer new nanoacoustic modes.  
Through a simple geometry, consisting of two identical coplanar elongated Au bars, we develop a comprehensive toolbox for the design of acoustoplasmonic metasurfaces for nonsymmetric hypersound transducers. To that end, we analyze the optical response to establish the coupling characteristics of the system and how symmetries are lifted. Afterwards, we study the nanoacoustic resonances to unveil the phonon dynamics in the case of systems with no plane-mirror symmetries. Then, we make analogies whenever the physical picture is similar in both systems and point out the main discrepancy points.  

\section{Description of the system and methodology}

 \begin{figure}[ht]
    \includegraphics[width=0.9\linewidth]{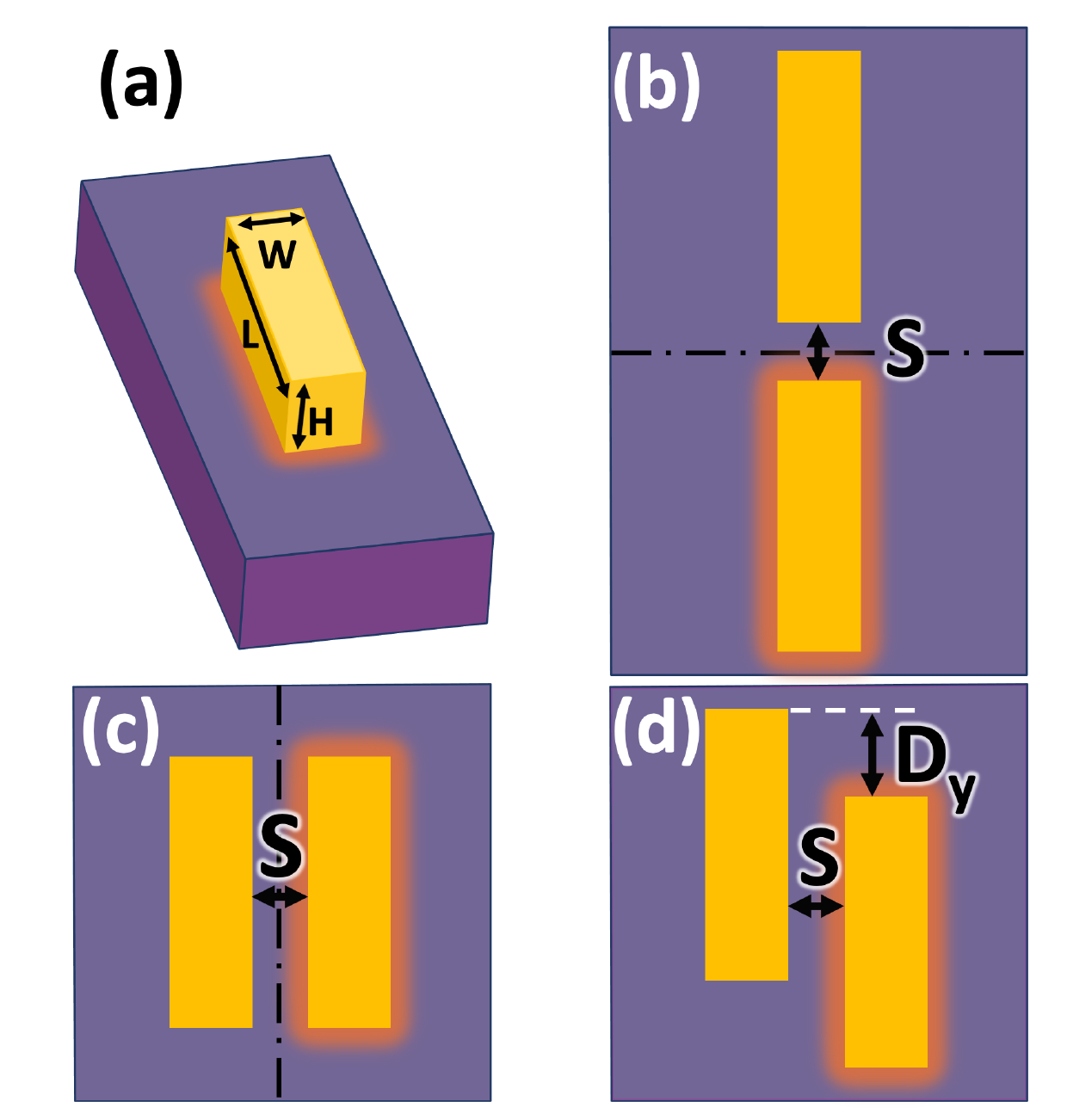}
    \caption{The basic system under study in this work consists on a Au bar (a) with dimensions L x W x H=100 x 30 x 30 nm$^3$. The interaction of two parallel co-planar bars on a silicon oxide, SiO$_2$, substrate is studied when they are spaced S=20 nm along the long-axis direction (b) or the short-axis direction (c and d). Additionally, a displacement D$_y$=75~nm along the long-axis direction is shown in (d) which breaks the mirror-symmetry of the system. A dark-orange illumination around the heated bar represents an initial 5K thermal excitation.}    
    \label{fig:system-geometry}
\end{figure}

The system under study consists of two gold bars with dimensions L=100 nm x W=30 nm x H=30nm (Length x Width x Height, as the one depicted in Figure~\ref{fig:system-geometry}(a)). To analyze the effects of symmetries, we explore the interactions of two co-planar bars on a SiO$_2$ substrate when they are separated S=20 nm along the long-axis direction, Figure~\ref{fig:system-geometry}(b), or short-axis direction, Figure~\ref{fig:system-geometry}(c). These two structures present clear mirror symmetries with respect to the planes depicted as doted-dashed lines in Figs. ~\ref{fig:system-geometry}(b,c).  A simple  displacement D$_y$ along the long axis (short axis) for the bars aligned along the short axis (long axis) will break the mirror symmetry, leaving only a rotational symmetry of 180 degrees, also known as C2-symmetry.

The Lumerical\textregistered~ software suite has been utilized to address the scattering and absorption cross sections. The computational solutions are based on the Finite Difference Time Domain (FDTD) method, with the dispersive optical constants for gold referenced in Ref.~\cite{Johnson_Christy:72}. A linearly polarized plane wave—incident along either the long or short axis of the nanostructure—with a uniform amplitude across the entire simulation cell, has been employed. The simulation cell dimensions (6 $\mu$m × 6 $\mu$m × 10 $\mu$m) were chosen to ensure that perfectly absorbing boundary conditions exert a negligible influence on the resultant electromagnetic fields. A highly refined mesh has been implemented within the nanostructures and the proximal near-field region (2.2 $\mu$m × 2.2 $\mu$m × 3.7 $\mu$m) with spatial resolutions of 1 nm × 1 nm × 1 nm (dx-dy-dz, respectively), which continuously expands to a maximum size of 10 nm as the distance from the near-field region increases towards the simulation boundaries, thus ensuring numerical convergence to the highest possible accuracy.

The Finite Element Method (FEM) in COMSOL\textregistered~ (Structural Mechanics Module) offers an effective framework to simulate the elastic response resulting from the isotropic thermal expansion in a metallic nanostructure.\cite{Bragas:23} The FEM simulation utilizes material parameters from COMSOL\textregistered's materials library. The nanobars are meshed with a maximum element size of 5~nm.  The SiO$_2$ substrate is a 1600 nm diameter semi-spherical region, surrounded by nonreflective domains (Perfectly Matched Layers, PML) that truncate the physical domain. To adequately represent the acoustic interactions between bars, a 300 nm diameter semi-spherical region of the substrate beneath the gold bars is meshed with a minimum element size of ~4 nm and a maximum element size of 50 nm.

The thermoelastic excitation of phonons is simulated by raising the temperature of a nanobar by a 5K initial thermal excitation, which is depicted by a dark-orange illumination around the heated bar in Figure \ref{fig:system-geometry}. The center of mass of each system, whether a single bar or double bars, is positioned at the origin of the coordinates. In studies involving two coupled bars, the geometric arrangement is designed to permit interactions (such as multiple wave scattering), both optical and acoustical. In order to allow counter-phase oscillations of the coupled bars,  thermal excitation is applied only to one of the bars.  Phonon detection is modeled by computing the integral of the root mean square (RMS) of the displacement on the two bars.

\section{Results}

\subsection{Optical response}

The optical properties of metallic bars or nanoantennae have been the subject of extensive research during the past decades. Recently, studies have focused on systems where two or more bars interact, leading to new optical phenomena \cite{Shifted_Femius_2016, BeaAcustoPlasm,Niek_doi:10.1126/science.1191922}. Despite the vast literature on plasmonics in interacting bars, \cite{Hybridizationmodel,Soukoulis_Chiral_Wang_2009,Shifted_Femius_2016,BeaAcustoPlasm,BeaSPIE} let us briefly describe the optical response of the structures depicted in Figure~\ref{fig:system-geometry}.

\begin{figure}[ht]
\includegraphics[width=0.9\linewidth]{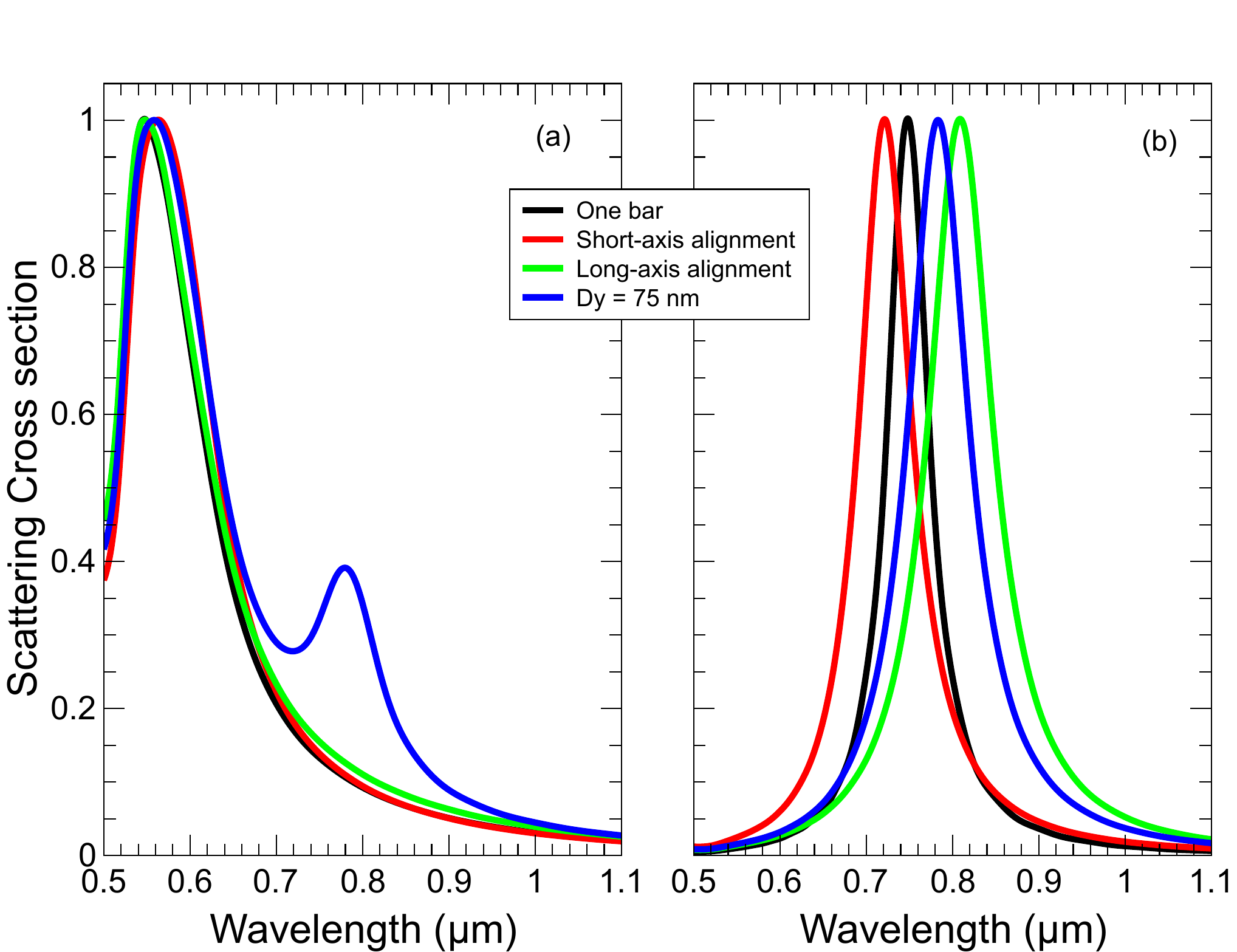}
        \caption{(a) Normalized scattering cross section for polarization along the short axis, depicting the single resonance at ca. 550~nm for all the mirror-symmetric cases (with the corresponding red and blue shifts due to interactions), and the simultaneous excitation of the long-axis resonance for the shifted case due to the broken mirror symmetry. (b) The same for polarization along the long axis. The resonance is ca. 750~nm and the blue and red shifts are more pronounced, but the weaker short axis resonance is withing the numerical noise level. 
        } 
        \label{fig:optics}
\end{figure}

In Figure~\ref{fig:optics} we present the normalized scattering cross section as a function of the wavelength of the incident light for the structures depicted in Figure\ref{fig:system-geometry} with D$_y$=75~nm. In Figure~\ref{fig:optics}(a) we present the results when the incident light is polarized along the short axis of the bars and in Figure~\ref{fig:optics}(b) when the polarization is along the long axis. The spectral location of the plasmon resonance for a single bar (or nanoparticle) is known to depend on the size of the bar aligned with the polarization direction, bigger size leads to longer resonance wavelength. That is nicely shown by the black line, which presents a resonance at ca.~550~nm for the polarization aligned along the shot axis and at ca.~750~nm when aligned along the long axis. This could be understood as one particle with an electric dipolar polarization with two resonances one for each principal axis. When two resonant particles are in interacting distance this picture is modified, and resembling molecular coupling, or coupled classical springs, the two resonances give rise to a coupled system whose resonances essentially depend on the separation.\cite{Hybridizationmodel,TwoBarsCortes2017} In our case, only one of the resulting hybridized resonances lie in the visible range of the spectrum, for both polarizations and separation distances along the short and long axis. For the two highly symmetric cases in Figure~\ref{fig:system-geometry}(b,c), where mirror symmetry is present, the interaction gives rise to red or blue shifts of the resonances with respect to the isolated bar (see red and blue lines in Figure~\ref{fig:optics}), but the equivalent hybridized electric dipoles would be aligned along the principal axis. However, when the mirror symmetry is broken, Figure~\ref{fig:system-geometry}(d), the equivalent hybridized dipoles would be misaligned with respect to the principal axis of the individual bars. That is equivalent to having a rotated dipole, and thus both resonances would be excited for both polarizations, along the short and long axis, but with different strengths. This is exactly the situation displayed in Figure~\ref{fig:optics}(a), where it can be seen that the resonance along the long axis is excited when the incident polarization is along the short one. It must be noted that the resonance corresponding to the short axis is also excited for polarization along the long axis, but the intensity is so low that it can barely be appreciated in Figure~\ref{fig:optics}(b) as it is comparable to the numerical noise level in our simulations.    

\subsection{Nanomechanical response}

\begin{figure}[ht]
    \includegraphics[width=0.9\linewidth]{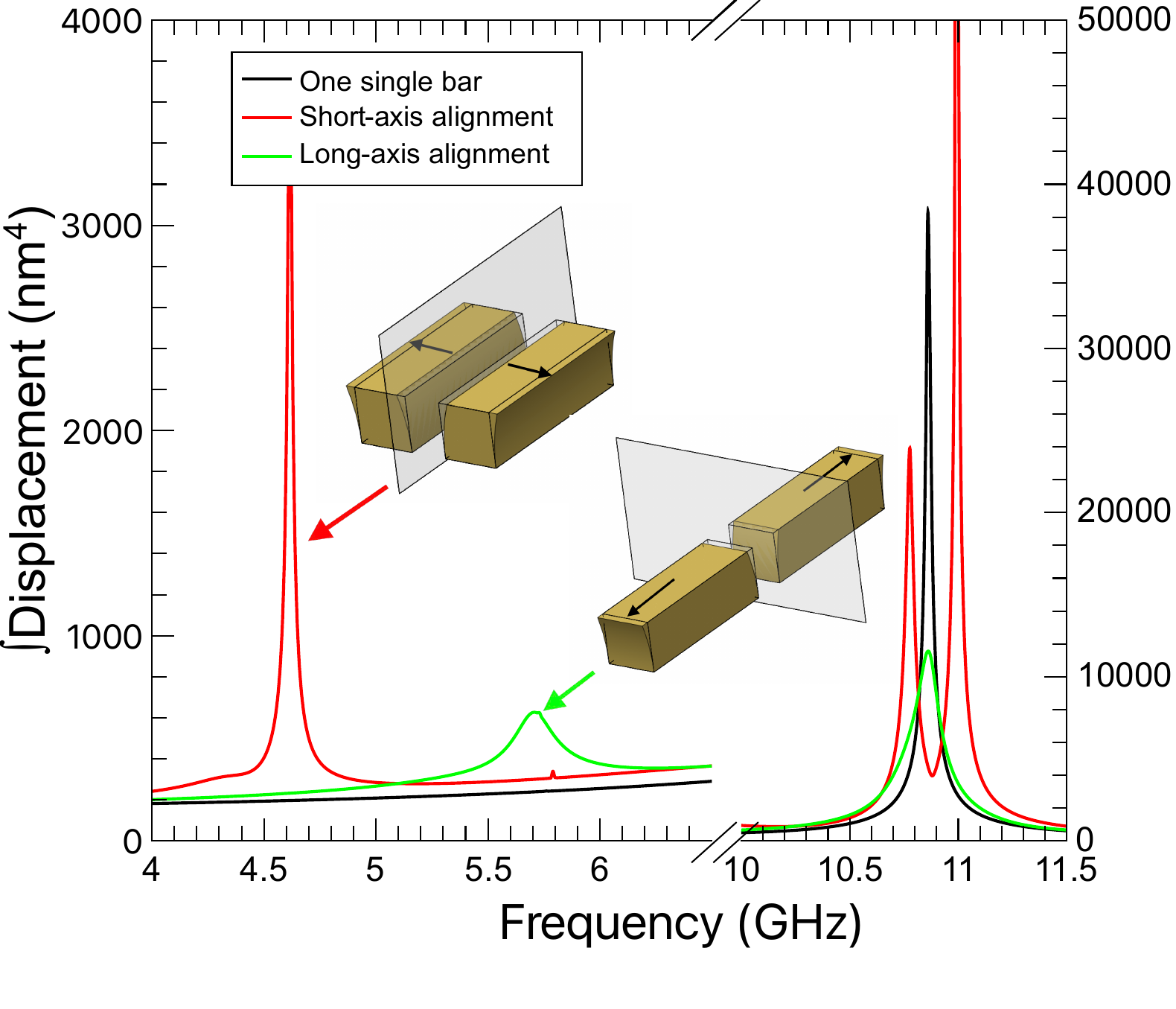}
    \caption{Integral of the RMS displacement as a function of frequency for the structures in  Figure~\ref{fig:system-geometry}(a-c). The insets depict the dynamics of the vibrational modes for the low-frequency region. Bars aligned along the short (long) axis show a vibrational mode schematically depicted by black arrows that correspond to a frequency peak marked with a red (green) arrow.}
    \label{fig:AcousticResSym}
\end{figure}

In order to unveil the physics introduced by the absence of a mirror symmetry, it is convenient to first understand the resonances associated to the most elementary systems presenting clear symmetry planes. When considering two nanobars, to find the allowed in-phase and counter-phase resonances for the interacting bars, we thermally excite only one of the bars (see Figure~\ref{fig:system-geometry}). 
We present in Figure\ref{fig:AcousticResSym} the resonances found in the high- and in the low-frequency regions (4 - 6.5 GHz -left- and 10 - 11.5 GHz -right-), for the mirror-symmetric structures. In the high-frequency region, the single bar presents only one resonance at 10.8 GHz corresponding to a longitudinal-like mode. However, the resonance at high frequency splits into two when the bars are aligned along the short axis: in-phase (lower frequency) and counter-phase (higher frequency) longitudinal modes. The alignment along the long axis leads to a single, in-phase, resonance at the same region.  

Let us now focus on the low-frequency range, where the single bar shows no resonances. In this region, two distinct peaks can be observed for the other symmetric alignments, along the short and long axes, at ca. 4.6~GHz and ca. 5.7~GHz. These peaks result from the interactions between the two bars through the substrate. The mode is essentially the same for both alignments, manifesting as low-frequency modes oscillating along the coupling direction, maintaining the appropriate mirror symmetry. 

\begin{figure}[ht]
    \includegraphics[width=0.9\linewidth]{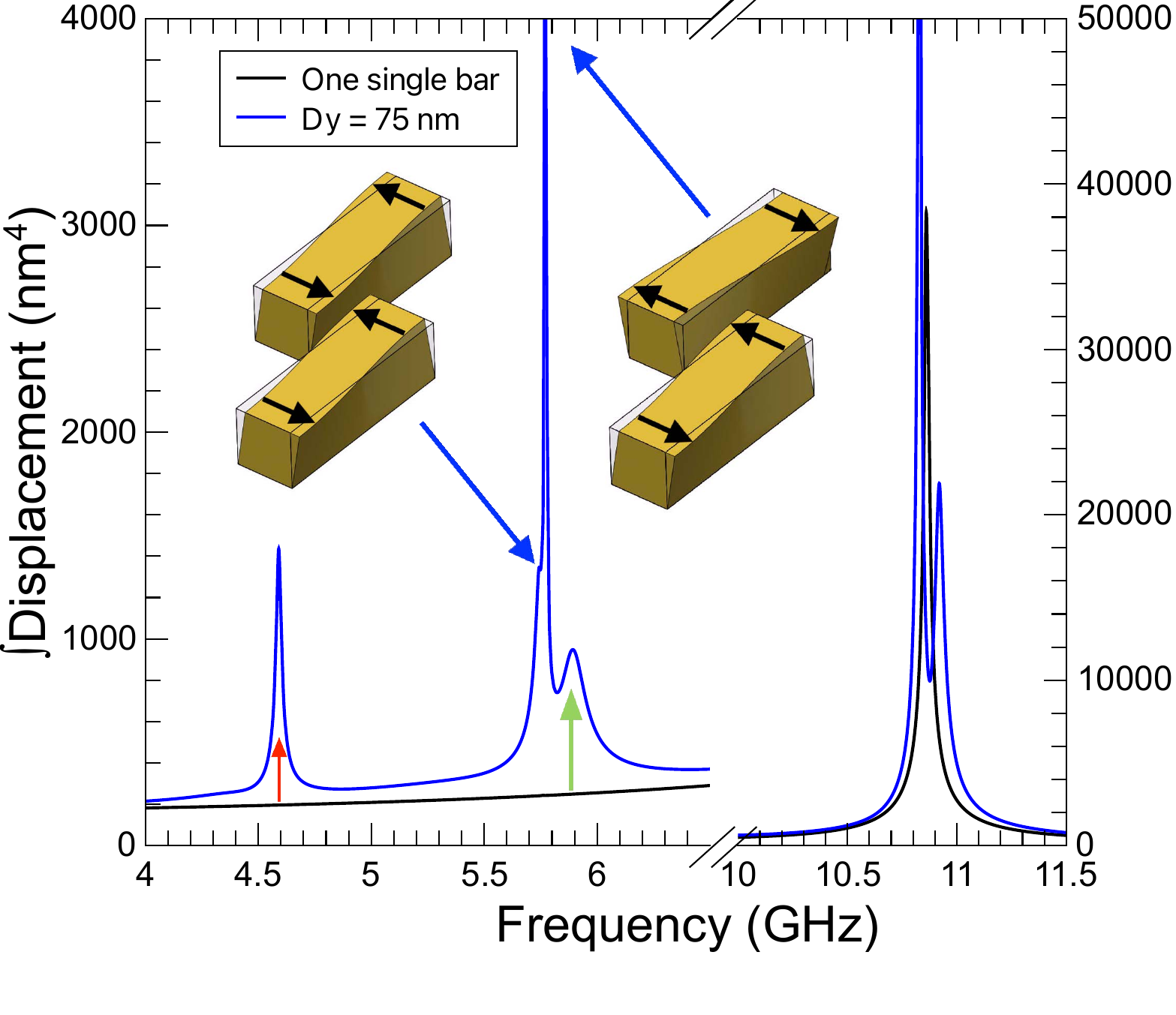}
    \caption{Integral of the RMS displacement as a function of frequency for the structures in Figure~\ref{fig:system-geometry}(d) for D$_y$=75 nm. The peak around 4.6 GHz (red arrow) has a similar origin as the one marked also with a red arrow in Figure~\ref{fig:AcousticResSym}. Blue arrows mark torsional vibrational modes schematically depicted in the corresponding insets and are the result of a crosstalking of horizontal and vertical alignment of the bars. Above these two resonances, there is another mechanical mode related to the peak marked with a green arrow in  Figure~\ref{fig:AcousticResSym}.}
    \label{fig:AcousticRes}
\end{figure}

Once the systems with mirror symmetries have been analyzed, we will consider the geometry depicted in Figure\ref{fig:system-geometry}(d), where the shift introduced by D$_y$ lifts the mirror symmetry and the only remaining one is a C2 in-plane rotational symmetry.
In Figure~\ref{fig:AcousticRes}, we show the simulation results for the structure with D$_y$ = 75~nm. 

The high-frequency range presents a behavior similar to the short-axis alignment, mirror-symmetric geometry, but with a smaller splitting. The mechanical mode profile of these two resonances is equivalent to the short-axis alignment. 

The low-frequency range presents a richer behavior. Similarly to the optical case, this geometry exhibits the two resonances present in the two mirror-symmetric ones, with a small frequency shift and are signaled with a red (green) arrow for the resonance associated to the short-axis (long-axis) alignment. The mechanical deformation profiles of these two resonances are equivalent to those schematically presented in Figure\ref{fig:AcousticResSym}, maintaining to a very high degree the mirror symmetries, and there is essentially no torsion in each of the bars. In contrast to the optical case, when mirror symmetries are lifted not only cross-coupling between long-axis and short-axis modes is allowed, but extra mechanical resonances at low frequencies are clearly visible at $\sim$5.7~GHz and $\sim$5.8~GHz. These new resonances display a mechanical deformation profile different to the previous ones. In contrast to the previous ones, the deformation is characterized by torsion. The two bars experience a torsion since, in both cases, the displacement of the ends of the bars is in the direction of the short axis and in opposite directions. The two resonances, as depicted as insets in Figure~\ref{fig:AcousticRes}, correspond to in-phase and counter-phase torsional deformations of each bar.

It is important to note that the ``counterphase'' resonances at $\sim$11~GHz in Figure\ref{fig:AcousticResSym} and at $\sim$5.8~GHz and in $\sim$10.8~GHz Figure\ref{fig:AcousticRes} are not allowed for a simultaneous initial thermal expansion of the two bars.

\section{Discussion and Conclusion}

We examined the effects of broken mirror symmetries in the optical and nanomechanical response of acoustoplasmonic resonators using a system composed of two identical elongated metallic bars. Our findings show that, the absence of mirror symmetries affects differently the optical and nanomechanical response. 

The analysis of the optical resonances through the optical cross-sections of  mirror-symmetric and  C2-symmetric coupled bars demonstrates that broken mirror symmetries induces a crosstalking between the resonances of the mirror-symmetric geometries. This is manifested by the excitation of resonances corresponding to the long axis when the light incidence is cross-polarized along the short axis.

In the nanomechanical case the situation is quite different. Interactions between bars produce a splitting of the nanomechanical resonances when aligned along the short axis. Additionally, new unconventional modes, preserving the mirror symmetry imposed by the symmetry appear in the low-frequency region. Broken mirror symmetries not only couple these nanomechanical modes  but introduces new resonant ones, whose deformation displays a distinct torsional profile.

To conclude, we show that broken mirror symmetries affect optical and nanomechanical resonances in a different way. Even in simple systems, such as the one considered here involving mirror-symmetric but interacting elements, it is possible to excite highly non-trivial resonances mediated by interactions.  These new nanomechanical modes present a marked torsional deformation profile. This deformation profile will eventually be transferred to the substrate, opening the way to the excitation of propagating acoustic modes with complex, tailored wavefronts. 

\subsection*{Disclosures}
The authors declare no conflict of interest.

\subsection* {Code, Data, and Materials Availability} 
The data and code supporting this study are available from the corresponding authors upon reasonable request.

\subsection* {Acknowledgments}
We acknowledge financial support from several founding agencies. BCLdL and AG-M acknowledge the Spanish Ministry of Science and Innovation (MCIN), AEI and FEDER (UE) through projects  PID2022-137569NB-C41 and TED2021-131417B-I00, and JGM acknowledges support through  project PID2021-124193OB-C22. NDL-K and CX acknowledge support from the European Research Council Consolidator Grant No.101045089 (T-Recs).

\bibliography{Stressors}   
\bibliographystyle{spiejour}

\end{spacing}
\end{document}